\documentclass{article}
\usepackage{arxiv}

\usepackage{iftex}
\ifPDFTeX

  \usepackage[utf8]{inputenc}
\else
  \usepackage{fontspec}
\fi
\normalfont

\usepackage{graphicx}
\usepackage{booktabs,array,longtable}
\usepackage{amsmath,amsfonts}
\usepackage{enumitem,placeins,etoolbox}
\usepackage{microtype}
\usepackage[english]{babel}
\usepackage{csquotes}
\usepackage[style=apa,backend=biber,natbib=true,sortcites=true]{biblatex}
\usepackage[hidelinks]{hyperref}
\ifPDFTeX
  \input{glyphtounicode}
\fi

\renewcommand{\headeright}{}
\renewcommand{\undertitle}{}
\renewcommand{\shorttitle}{Network Analysis in Communication Research}
\renewenvironment{abstract}
  {\centerline{\large\bfseries Abstract}\begin{quote}}
  {\end{quote}}
\AtBeginBibliography{\raggedright}

\makeatletter
\patchcmd{\@maketitle}{\textsc{\undertitle}\\}{}{}
  {\PackageError{Preprint}{Unexpected arxiv title layout}{Check arxiv.sty.}}
\patchcmd{\@maketitle}
  {\vskip 0.4in \@minus 0.1in \center{\@date} \vskip 0.2in}
  {\vskip 0.2in}{}
  {\PackageError{Preprint}{Unexpected arxiv date layout}{Check arxiv.sty.}}
\makeatother

\newcommand{\figref}[1]{Figure~\ref{#1}}
\newcommand{\tabref}[1]{Table~\ref{#1}}
\newcommand{\secref}[1]{Section~\ref{#1}}
\newcommand{\appref}[1]{Appendix~\ref{#1}}

\newcommand{\NetworkEligibleRecords}{2{,}103}
\newcommand{\NetworkPrimaryNodes}{30}

\hypersetup{
  pdftitle={Network Analysis in Communication Research: A Review of Methods, Theories, and Research Practices},
  pdfkeywords={network analysis, communication research, integrative review, methods, theory, research practices}
}

\title{Network Analysis in Communication Research:\\A Review of Methods, Theories, and Research Practices}

\author{Pengjia Cui\\
  Computational Social Science, School of Social Sciences\\
  University of California San Diego\\
  \href{mailto:pcui@ucsd.edu}{\texttt{pcui@ucsd.edu}}\\
  ORCID: \href{https://orcid.org/0009-0004-6360-8897}{0009-0004-6360-8897}
}
\date{}

\begin{document}
\maketitle

\begin{abstract}
Network analysis has become a standard way of representing consultation, sourcing, media use, attention, and influence in communication research. This review examines 31 applications across communication subfields and asks of each what communication problem it addresses, which explanatory resource it invokes, and what its research practice makes observable. The three dimensions form a comparison framework and yield three propositions with stated scope and refutation conditions: resource relevance is task- and criterion-dependent; commonality at one level of aggregation does not imply commonality at another; and influence estimates are conditional on relationship type, group stage, and time. A short bibliometric section describes how the literature's cited references are organised: method references are paired with one another more often than year-specific margins predict (ratio 1.36) and with theoretical references less often (0.82), a pattern consistent with a methodological toolkit that circulates apart from the commitments that license its use. The review closes with designs that would test each proposition.
\end{abstract}


\section{Introduction}

Communication network research asks how people obtain knowledge, why shared attention divides into different interpretations, and under what conditions relationships shape what people believe and use. Its studies represent consultation, friendship, media use, citation, and lexical association as edges. Because the vocabulary is shared, a reader can move between these studies as though they were asking the same question in different settings.

This review asks whether they are. Its argument is that a shared vocabulary conceals three differences that determine whether findings can be placed side by side: what a study's nodes are, what its edges are taken to mean, and what it treats as needing explanation. Two studies that both report centrality can be describing recognition of expertise, shared media attendance, or co-occurrence of words, and the number carries a different claim in each case. Centrality measures are not interchangeable even within a single formalism \citep{freemanCentralitySocialNetworks1978a}; the same edge can stand for friendship, animosity, professional acquaintance, or geographical proximity \citep[p.~172]{newmanStructureFunctionComplex2003a}; and what a network representation licenses depends on the abstraction that produced it \citep{brandesWhatNetworkScience2013,buttsRevisitingFoundationsNetwork2009}. Comparison therefore cannot proceed by matching measures.

To state the problem, the review borrows one distinction from network science. \citet[pp.~3--5]{brandesWhatNetworkScience2013} distinguish two steps in any network model: a phenomenon is first abstracted into a network concept, and that concept is then represented in data. Interpretation reverses both steps, so that substantive theory is required to carry a statement about a network representation back to the phenomenon; abstraction facilitates comparison and, by the same token, obstructs that return. \citet{buttsRevisitingFoundationsNetwork2009} makes the corresponding measurement argument: the definition of vertices, edges, and time scale is a substantive choice, and conclusions drawn from a network can be artefacts of that choice rather than properties of the system. Social-network theory adds that ties of different kinds---similarities, social relations, interactions, and flows---support different mechanisms \citep{borgattiNetworkAnalysisSocial2009}, and that explaining outcomes through networks is a different enterprise from explaining networks themselves \citep{borgattiNetworkTheory2011}. These arguments are general. What they do not supply is a procedure, specific to a substantive field, for deciding when two empirical findings rest on the same abstraction and the same representation.

Communication research has reviewed its network studies in two ways that stop short of that procedure. \citet{shumateTaxonomyCommunicationNetworks2013} distinguish flow, affinity, representational, and semantic relations, and synthesise the antecedents, outcomes, and processes associated with each; grouping studies by relation type is itself a way of comparing findings within a type. \citet{fuAreWeMoving2020} map the intellectual structure of online social network research through bibliographic coupling, co-word, and co-citation analysis, and call for closer integration of network theory with the properties of the networks analysed. Neither treats the practice that constructs a network---nomination, affiliation, co-occurrence---as a separate axis of comparison, and neither specifies when a difference between two studies is a disagreement about the world rather than a difference in what was observed.

This review sets out that procedure for communication research. It is an integrative review in the selective sense: it compares a purposively selected set of applications rather than screening an eligible population, and it builds a framework from those applications rather than estimating a distribution \citep{torracoWritingIntegrativeReviews2005,cooperOrganizingKnowledgeSyntheses1988}. Selecting cases to maximise contrast, rather than to represent the population, is the present review's design and not one of the coverage categories those sources define. Its subject is the application of network analysis in communication research---the methods studies use, the theories they invoke, and the practices through which they construct a network.

The primary contribution is a comparison framework built from the reviewed applications. It separates three dimensions that can be read off a published study: the communication problem a study addresses, which fixes the phenomenon; the explanatory resource it invokes, which licenses the abstraction and the return from data to phenomenon; and the research practice through which it constructs a network, which fixes the representation. The dimensions vary independently, which is the point. Two studies may address the same problem with different resources, or the same resource with different practices, and in either case a difference in findings is not yet a disagreement. Applying the framework to 31 applications yields three propositions, each with a stated scope and a stated refutation condition (\secref{sec:propositions}). The propositions are substantive claims about communication---about relevance, shared attention, and influence---rather than restatements of the measurement argument, and each identifies a design the reviewed literature lacks.

A short bibliometric section then turns from the applications to the literature that contains them. It asks whether the field's cited references join explanatory and methodological resources, and finds that they largely do not: method references are paired with one another above expectation and with theoretical references below it (\secref{sec:diagnostic}). The section describes citation structure; the problem-level test that would connect it to communication problems remains open (\secref{sec:agenda}). The full analysis is in \appref{app:cocitation}.

The review and the bibliometric section bear on each other. A knowledge base in which methods travel together and apart from theory admits studies that share a measure without sharing the commitments that make their findings commensurable. That is a structural account of why a methodologically sophisticated field can accumulate procedures while its claims remain difficult to compare, and it is an account the review states in a form that can be tested.

The argument proceeds in the order in which the framework is used. \secref{sec:framework} sets out the three dimensions and the comparability rule they imply. \secref{sec:corpus} describes the corpus and the reading procedure. \secref{sec:relevance} to \secref{sec:influence} develop three comparisons: resource relevance, shared attention and differentiated meaning, and influence across relationships and periods. \secref{sec:propositions} states the propositions and their refutation conditions. \secref{sec:extensions} tests the boundaries of the account with cases on support, collective action, and representation. \secref{sec:practices} surveys the five constructions that recur across the corpus and what each makes observable. \secref{sec:theories} places the applications within the theoretical families that supply their explanatory resources. \secref{sec:diagnostic} adds a bibliometric section on the literature's citation structure, with the full analysis in \appref{app:cocitation}. \secref{sec:agenda} specifies the designs that would settle what remains open. Appendices hold the retrieval record, the co-citation specifications, the case roster, and the observation guide.

\section{A framework for comparing network studies}
\label{sec:framework}

\subsection{Three dimensions}

The framework treats three properties of a study as independent, in the sense that fixing one does not fix the others. Each corresponds to an element of the network-model sequence described in the introduction. The communication problem is the phenomenon. The explanatory resource is the substantive theory that justifies abstracting that phenomenon into nodes and ties and that is needed again when results are interpreted. The research practice is the representation: the observational procedure that turns the network concept into data. What the framework adds to the general sequence is that, in communication research, each element can be identified from a published study and compared across studies.

\paragraph{Communication problem.} What the study needs to explain. Four families recur across the corpus: how contact and coordination develop across disagreement; whom people attend to and how relationships and media use change together; how issues, actors, and evaluations become associated; and how expertise, sources, and media become accessible. These families overlap, and one study can address more than one. The groupings describe what a study is trying to explain, not which subfield it belongs to.

\paragraph{Explanatory resource.} What the study invokes to explain the problem, and therefore what a tie is taken to do. Transactive memory makes recognised expertise relevant to consultation. Institutional sourcing connects available sources to reporting routines. Repertoire and audience-duplication accounts distinguish combinations of selected content from shared attendance. Agenda setting motivates correspondence between the salience of elements; framing directs attention to evaluative targets and attributed responsibility. Homophily and co-orientation require separating tie selection, shared preference, and change within an existing tie. These resources are not interchangeable labels. Each specifies a different mechanism---a tie as a channel through which something flows, as a marker of similarity, or as a relation that coordinates---and therefore a different observation that would count as evidence.

\begin{figure}[htbp]
\centering
\includegraphics[width=\textwidth]{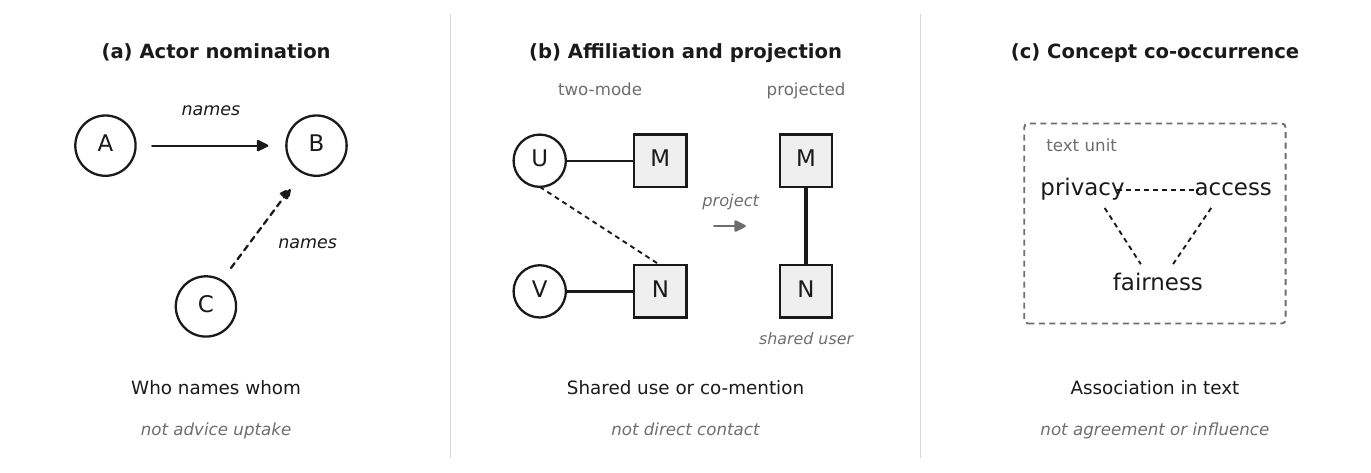}
\caption{Three edge constructions and what each supports. (a) Reported nominations identify whom an actor names, without establishing that advice was taken. (b) Two-mode affiliation between users or posts (U, V) and outlets or sources (M, N), with the open arrow marking projection: an edge between outlets records a shared user, not direct contact. (c) Co-occurrence within text units records that elements appear together, not that they are agreed with or causally linked. Weights and study-specific thresholds are omitted; the panels are schematic.}
\label{fig:relation-guide}
\end{figure}

\paragraph{Research practice.} What the study's network makes observable. Nominations identify reported partners; shared affiliations reveal overlapping participation and, when projected, create links between outlets or sources that share users; textual co-occurrence shows which elements of meaning appear together within a unit of text; longitudinal ties record who names whom at successive occasions. These constructions support different inferences and foreclose others (\figref{fig:relation-guide}). A projected affiliation edge does not record contact. A co-occurrence edge does not record agreement. A nomination does not record use.

\subsection{The comparability rule}

The three dimensions yield a rule that is simple to state and demanding to apply.

\begin{quote}
Two findings may be juxtaposed when they address the same problem, invoke the same explanatory resource, and rest on practices that make the same observation. Where they differ on a dimension but not on the others, the difference may be informative but is not yet a disagreement. Where they differ on the dimension that the claim depends on, the claim is not comparable and neither supportive nor contradictory evidence.
\end{quote}

The rule is deliberately conservative. It does not say that cross-dimension comparison is impossible; it says that such comparison must first establish what each study makes observable. Most of the comparisons below are of the second and third kinds: studies addressing the same problem with different practices, or with different resources, whose findings can be placed side by side once the difference in practice is specified.

Two consequences follow. First, compatibility is not the same as agreement. Two studies can be compatible because they observe different levels of the same process, and this is a weaker relation than replication. Second, an unresolved contrast is a result rather than a failure. Where designs vary on several dimensions at once, the contrast identifies conditions that any explanation must distinguish, and that is a claim about what the evidence can currently support.

\subsection{What the framework does not do}

The framework locates a study's claim relative to other claims and says what further evidence would settle a difference. It does not estimate the prevalence of a practice, rank studies by quality, adjudicate between theories, or propose a formal theory of networks; the mathematical properties of representations are taken as given. These limits are why the framework can be applied to a purposively selected corpus rather than to a screened population.

\section{Corpus and reading procedure}
\label{sec:corpus}

\subsection{Candidate corpus and retrieval boundary}

The candidate corpus is 2,114 Web of Science Core Collection records retrieved from five Plain Text exports dated 31 August 2026. Retrieval combined the Communication category with network-analysis expressions, network objects, models, and structural indicators, followed by Article or Review document-type filtering. Query Q-R was confirmed by the author as the expression used for the final export; the execution history and complete export settings have not been recovered. The Topic field searches titles, abstracts, author keywords, and Keywords Plus, so records could be retrieved through indexing terminology as well as authors' descriptions. \appref{app:retrieval} reproduces the expression and records its provenance.

The corpus is bounded by the Communication category. Original document-type strings are preserved, including 1,905 records typed \emph{Article}, 28 typed \emph{Review}, and composite types such as \emph{Article; Early Access} and \emph{Article; Proceedings Paper}. The pool supports literature-level description; it has not been individually screened into a census of eligible studies, and its size is not the number of studies synthesised.

\subsection{Eligibility}

A study enters substantive comparison when it uses an identifiable relation structure, network matrix, network model, or networked text to address a communication question. A network metaphor, a platform setting, or a technical use of network terminology alone does not meet that criterion. Foundational and contextual sources outside the pool are identified separately and are not counted toward the corpus.

\subsection{Selection and its limits}

Studies were selected purposively to carry, qualify, or bound the comparisons, subject to access to verifiable texts. The 31 applications comprise 13 core cases that carry the principal comparisons, six supporting cases that contextualise or qualify them, and 12 extended cases that test the account's boundaries; one further application is developed in \appref{app:roster}. Discovery routes included a 40-record corpus audit, a 24-case full-text reading, a 64-record co-citing expansion, and the supplementary searches described below; these sets are purposive and overlapping, not stages of a screening flow. The initial selection order cannot be fully reconstructed. Supplementary Data S1 records each study's role, inclusion rationale, evidence level, and source locator.

Four features of the selection limit what the review can claim, and they are stated at the outset rather than in the limitations. First, the corpus was not screened into an eligible population, so the 31 applications are not a sample from which prevalence can be estimated. Second, purposive selection targets cases that produce contrast, which means the review over-represents studies whose findings do not agree. Third, abstract-level evidence is used where full texts were unavailable, and every such use is identified in the text. Fourth, three applications were removed from an earlier roster of 33 on grounds of citation impact and peripheral contribution; one of them, \citet{fuhseAnalyzingNetworksCommunication2023}, is reported as a supplementary application in \appref{app:roster}, and the other two do not appear in the manuscript. A review that uses citation structure as diagnostic evidence should note that citation impact also entered the case-selection step.

\subsection{Reading procedure}

For each comparison the review records the communication question, the explanatory resource, the relation definition, the population, the observation period, and the inferential limits, and interprets reported findings against these. Studies provide complementary evidence when they illuminate distinct aspects of a shared question. Findings are compatible when they concern different levels or outcomes and can hold together. A contrast remains unresolved when designs vary in ways that prevent attribution to a particular explanation. Claims about relation construction, sampling, timing, and model specification draw on full texts.

Citation functions were examined at page level in two cases, the BP crisis study and the health-journalist study. Independent duplicate coding and an intercoder-reliability estimate are not documented; the author performed the selection and coding, and the absence of a second coder is a limitation rather than a procedural detail.

A supplementary search on 19 September 2026 screened 171 unique records for evidence that could qualify the influence, sourcing, and media--public comparisons. A full-text queue capped at two records and ordered by database identifier (accession order, which is reproducible but does not rank by substantive relevance) added two agenda studies; records above the cap were not inspected. A subsequent search on 22 September added three agenda studies bearing on affective correspondence, political predispositions, and shared commenters. These extensions covered agenda research more thoroughly than the other comparisons, and the asymmetry is visible in \secref{sec:attention}.

\subsection{Large language models}

Large language models assisted literature screening and coding, comparison of source texts, and the development and checking of analysis and visualisation code. Complete historical records of model versions and prompts are unavailable. Supplementary Data S1 and the archived diagnostics identify the documented source and numerical checks and the limits of coding verification. Because no intercoder reliability was estimated, the LLM assistance cannot be described as independent verification; it is a documented aid whose outputs were checked against source texts. The archived code also records two model-assisted readings of the same 24 case abstracts, written independently and then adjudicated by the author, for the problem--family coding that a problem-level test would require (\secref{sec:agenda}); that coding is not used in the analyses reported here, and agreement between two model readings is not human intercoder reliability.

\section{Comparison one: what makes a resource relevant}
\label{sec:relevance}

\subsection{Recognition, importance, and use are different criteria}

Transactive memory explains organisational knowledge seeking through recognition of who knows what. \citet{palazzoloOrganizingInformationRetrieval2005} surveyed 154 members of 12 pre-existing teams, separating self-rated expertise, others' recognition of expertise, and topic-specific consultation nominations. Perceived expertise aligned with anticipated consultation; self-rated expertise was a weak predictor. Exponential random graph models conditioning on the number of ties and controlling for reciprocity and transitive closure also indicated that members tend to have more than one consultation tie, so accounts of uniformly specialised, one-way retrieval describe the theory better than they describe the teams. Configurations involving several members cannot be captured at the individual level of analysis alone \citep[p.~748]{palazzoloOrganizingInformationRetrieval2005}.

The distinction is consequential because having knowledge and being sought for it rest on different judgements. A person can hold relevant knowledge that no colleague recognises, and a person can be consulted for reasons other than expertise. Neither outcome is a proxy for the other.

\citet{taylorBuildingInterorganizationalRelationships2003} studied a different criterion in a different setting: 17 Croatian civil society organisations that provided nonpartisan information during the 2000 parliamentary election campaign, with 13 of the 17 organisations responding. Representatives reported the communication channels used with each partner, yielding a multiplex index that combines the number and the assigned richness of channels, and separately rated the value of the communication relationship with each organisation on a 0--10 scale. An open-ended question, asked before respondents saw the roster, elicited nominations of the organisations most important to the campaign. The channel index correlated strongly with relationship-importance ratings (quadratic assignment procedure $r=.78$, $p<.001$). Nomination and centrality diverged for the most central organisation: the international donor USAID had the highest multiplex in-degree and was never named in response to the open-ended question about organisations that mattered most to the campaign, while GONG, second in multiplex in-degree, was named most often (33 nominations). Across the 17 organisations the correlation between multiplex in-degree and campaign-importance nominations was $r=.52$ ($p<.05$).

Contact and attributed value therefore align closely on one criterion and only partly on another. The finding is not that one measure is wrong. It is that a partner's importance to a continuing communication relationship is a different quantity from its importance to a collective outcome, and a study that measures only one cannot speak to the other.

\subsection{Journalistic sourcing separates monitoring from use}

Journalism research supplies a third criterion and the mechanism that connects them. \citet[pp.~256--261]{tuchmanNewsNet1978} locates source availability in the deployment of reporters across territories, institutions, and topics, and newsworthiness in organisational routines and editorial negotiation. The net of placements favours some accounts of social reality over others before any individual story is written. \citet[pp.~223--224]{zelizerJournalistsInterpretiveCommunities1993} adds that journalists form an interpretive community through the circulation and negotiation of shared discourse, which matters here because dense peer connections alone would not demonstrate that circulation.

Following networks observe positions within this environment. In \citet[pp.~685--687]{verweijTwitterLinksPoliticians2012}, several reporters of the Dutch public broadcaster appear among the most extensive followers of political accounts without appearing among the most followed, while an RTL journalist leads the incoming-following ranking. Seeking access and attracting attention are different positions in the same profession, and the direction of the tie is what makes them visible.

The task-based explanation comes from interviews. \citet{johnsonMuchAdoNothing2018} combined following networks of 33 Flemish economic journalists, collected between April and August 2015, with retrospective interviews conducted between May and October 2016. Interviewees recalled the sources used in the week preceding the interview. Major source groups recurred across the two records, but their shares differed substantially: media and journalists accounted for 28\% of followed accounts but 21\% of reported sources, business professionals for 21\% and 37\%, and experts and academics for 9\% and 17\%. Interviewees described Twitter as a source of story ideas and trend monitoring that could prompt a telephone call, and distinguished background use from explicit citation in published news \citep[pp.~875--882]{johnsonMuchAdoNothing2018}.

The recurrence of major source groups shows continuity between access and use. The differences in share show that a following list is not a specification of a sourcing repertoire. The interviews give the difference a mechanism: media peers support monitoring, while business professionals and experts support elaboration and a distinctive angle. Monitoring and task use are different roles rather than points on a single scale of access.

Two limits bound the inference. The comparison pairs retrospective accounts of one workweek with following data collected roughly a year earlier, so recall and intervening change may explain part of the divergence. And the interviews do not track a continuous production sequence or establish which sources appeared in published articles.

\subsection{Citizens as sources and the criterion of representation}

A fourth criterion enters when the resource is not expertise but experience. Some interviewees distinguished contact with readers from sourcing: citizen complaints occasionally supplied leads but often did not develop into publishable accounts \citep[p.~881]{johnsonMuchAdoNothing2018}. An audience member can therefore have contact with a journalist without being treated as an informant.

This bears on an argument that \citet[pp.~1662--1663]{zhangHealthJournalistsSocial2024} review rather than establish: that ordinary citizens contribute experiences and perspectives not supplied by professional experts, and that social media can broaden access to them. Their own analysis of a co-@mention network among health journalists during the early COVID-19 emergency found the opposite pattern: source repertoires concentrated on elite actors, especially news media outlets, health journalists, and public health officials, with citizens marginalised. The finding is a source concentration result, and the argument about citizens' distinctive contribution is a reason to worry about that concentration, not evidence against it.

\subsection{What the comparison establishes}

Recognition, relationship importance, campaign contribution, and reported source use are four criteria for judging a resource's value. They are not four measures of one quantity. The sourcing interviews explain why the criteria diverge: a relationship is valuable for what it permits in a particular task, and tasks differ.

The comparison illustrates \textbf{P1}: for a given resource, judgements of relevance vary by task and criterion, and recognition, attributed importance, and reported use need not converge. The scope is settings in which actors purposively select sources, and the claim would fail if a single task-independent criterion predicted all three outcomes equally well across tasks. No reviewed study tests that possibility.

\section{Comparison two: shared attention and differentiated meaning}
\label{sec:attention}

\subsection{Two levels of aggregation}

Audience networks connect outlets through shared users. They reveal common points of attention that separate audience totals miss, and they do so even when outlets differ greatly in reach. What they do not reveal is what people selected within the outlets they share, and that gap is where the comparison begins.

\citet{websterDynamicsAudienceFragmentation2012} situate choice within interdependent providers, users, and measurement systems. In March 2009 panel data covering 98 television channels and 138 internet brands with a reach of at least 3\%, extensive duplication coexisted with unequal outlet popularity. Duplication describes overlapping use, not interpersonal contact, and a media-centric account of outlet size is not interchangeable with a user-centric account of repertoires or an audience-centric account of shared use across outlets.

\citet{rauchfleischTransnationalNewsSharing2020} supply the article-level comparison. Combining URL sharing with followee structures, they identified media repertoires and audience communities for Swiss news media. The German-speaking Swiss community and the German far-right community both shared articles from the \emph{Neue Zürcher Zeitung}. The most-shared articles differed: a broad range of domestic and scientific topics in the first, and Germany, Chancellor Merkel, Muslims, and refugees in the second \citep[pp.~1223--1224]{rauchfleischTransnationalNewsSharing2020}. A common outlet does not imply a common article repertoire.

Two applications from outside the candidate pool extend the outlet-level comparison to the role of algorithmic curation and ideological segregation in attention formation. \citet{bakshyExposureIdeologicallyDiverse2015a} analysed 10.1 million United States Facebook users over seven days in January 2015. Ideological homophily was measured in the friend network; exposure measured potential contacts from friends' sharing activity; and individual click-through was the observed act. Comparing the three levels---network structure, algorithmic News Feed ranking, and individual choice---the study found that algorithmic ranking reduced cross-cutting exposure by an estimated 8\% for liberals and 5\% for conservatives, while individual choice was the stronger limiter. The finding turns on all three levels being separable: a study measuring only following networks would not distinguish the structurally available from the algorithmically presented or the individually clicked.

\citet{barberaTweetingLeftRight2015} estimated ideological positions for 3.8 million Twitter users and mapped co-following and retweet networks for twelve political and nonpolitical issues. For political issues such as the 2012 presidential election and the 2013 government shutdown, information was exchanged primarily among ideologically similar users. For nonpolitical issues such as the 2014 Super Bowl, exchanges were not sorted by ideology. The Newtown shootings began as cross-ideological exchange before sorting over time. The findings locate P2's level-dependence in both issue type and temporal stage: within one network, the level at which segregation appears depends on what is being discussed and when it is being discussed.

This is the clearest instance of P2 in the corpus: commonality at the outlet level coexists with differentiation at the article level. The observed act is sharing, not verified reading or endorsement, and the comparison is between two communities rather than between the same users over time.

\subsection{Audience composition as a basis for choice}

A different question is whether the perceived composition of an audience affects what is chosen. \citet{dvir-gvirsmanMediaAudienceHomophily2017} combined a panel that constructed participant networks from repeated visits to the same websites, yielding individual external--internal scores that compare connections to people with opposing versus shared political orientations, with a randomised experiment that varied audience-identity cues alongside the stated political orientation of content. Audience cues changed content choice. The conditions did not differ significantly in self-placement, but participants given congruent content-and-audience cues perceived themselves as more extreme than those given content information alone.

The polarisation finding therefore depends on the outcome measured, and the network measure and the experimental manipulation address distinct parts of the proposed process. The design supports an account in which audience identity is one basis for choice; it does not establish that homogeneous audiences produce polarisation.

Composition also has an aggregation problem. \citet{ruscheFewVoicesStrong2024} examined German Bundestag members' followers in October 2018, averaging each politician's followers' shares of connections to that politician's party within the sampled legislators rather than across all accounts those followers follow. A user who follows many politicians contributes to several audiences. For the Alternative for Germany, a strongly committed group made up about 7\% of the party's unique followers but 55--75\% of most of its politicians' follower audiences. A small group can be disproportionately present across many audiences without constituting most users.

\subsection{Three routes to apparent media--public disagreement}

The agenda-setting tradition supplies the third comparison. \citet{mccombsAGENDASETTINGFUNCTIONMASS1972} established correspondence between media emphasis and the importance voters attach to issues. Network agenda setting extends the question from elements to the relations among them: repeated joint presentation can strengthen joint retrieval, alongside prior schemas and active elaboration \citep{mccombsExpandedPerspectiveAgendaSetting2012}. \citet[pp.~52--53]{entmanFramingClarificationFractured1993} defines framing through problem definition, causal interpretation, moral evaluation, and treatment recommendation, and distinguishes frames in communicators, texts, receivers, and culture, with the explicit caution that textual frames do not guarantee audience effects. The traditions overlap in studying associations among attributes, and framing is often treated as a refined form of second-order agenda setting \citep[p.~98]{schultzStrategicFramingBP2012}.

Baseline correspondence is positive. \citet[pp.~674--679]{vuExploringWorldOutside2014} compared United States news issue networks with Gallup poll data from 2007 to 2011, with ties counting issue pairs appearing among the top five categories. Annual media--public correlations ranged from $r=.65$ to $.87$; six of nine cross-lagged comparisons favoured the media-to-public direction, two the reverse, and one was reciprocal. The public networks aggregate issue co-prominence rather than associations elicited from individuals, and common events remain an alternative explanation.

The first route to divergence is that correspondence in issues can coexist with difference in evaluation. \citet[pp.~304--311]{vargoNetworkIssueAgendas2014} linked issues mentioned by the same Twitter user or media source on the same day and aggregated by week during the 2012 United States election. Supporters' agendas aligned with several media networks, including opposing-party media; yet models of issue centrality fitted Obama supporters best with general-news media and Romney supporters with Republican-oriented media. Correspondence and selective alignment coexist without observed exposure or within-person change.

Predispositions are already internal to this programme. \citet{melekPredispositions2024} compared 1,003 news items with a 1,200-person cognitive-map survey on Syrian refugees in Türkiye. Pro-government and opposition media networks correlated at $r=.96$. Pro-government media and pro-government respondents corresponded at $r=.73$ ($p<.05$), while the opposition pairing was not significant. The pattern does not establish a difference between groups, and issue similarity is not agreement: six broad themes can organise shared concern while concealing opposed attributions.

Affective attributes extend the account and expose its measurement dependence. \citet{liuAffectiveNAS2026} compared Chinese newspaper coverage of China--United States relations with a 514-person mind-mapping survey informed by focus groups, distinguishing substantive from affective attributes. Raw-count QAP analyses showed positive correspondence for substantive, positive-affect, and negative-affect networks; cosine-normalised analyses produced different patterns. Conclusions about affective correspondence therefore depend on whether attribute prevalence is held constant, and valence-coded association is an object of agenda correspondence rather than evidence that evaluation belongs only to framing.

The second route is that a shared issue can carry a reworked evaluation. \citet{hsiaoNetworkAgendaSetting2026} compared right-wing United States YouTube news with its comments. Shared issues appeared alongside stronger partisan and negative associations in the comments, and a shared evaluative term could change its target: criticism of pandemic modelling became a claim that the pandemic itself was a hoax. Election coverage showed more overlap between media and comments than pandemic or Black Lives Matter coverage. The result qualifies a strong copying expectation through differentiated reorganisation; it does not negate correspondence where it occurs, nor establish that particular users changed their interpretation.

The third route is that different people participate. \citet{luNetworkedFraming2024} analysed 643 vaccine-crisis videos and 26,852 comments. Comment topics did not differ significantly across episodic, thematic, and off-topic video-title frames, though the descriptive proportions varied. A shared-commenter network associated overlap with channel stance and subscription relations. Divergence between media frames and comment discourse therefore need not be reinterpretation by the same participants; it can reflect who gathers where.

\subsection{What the comparison establishes}

Three routes to apparent media--public disagreement are distinguishable: shared issues carrying different affective associations or evaluative targets; groups aligning selectively with media; and different participants producing different discourse. Network agenda setting contributes evidence on substantive and affective correspondence, framing on the organisation of responsibility and evaluation, and audience research on participant selection.

The comparison illustrates \textbf{P2}: commonality at one level of aggregation does not imply commonality at another. Outlet overlap does not imply article agreement; issue correspondence does not imply evaluative agreement; shared topics do not imply shared participants. The claim would fail if outlet overlap predicted article-level agreement, or issue correspondence predicted evaluative agreement, within the same participants. No reviewed study measures both levels on the same people, and the agenda studies in particular link aggregate networks rather than individuals.

\section{Comparison three: influence across relationships and periods}
\label{sec:influence}

\subsection{Three processes that produce the same observation}

Longitudinal network models distinguish the formation of ties among people who are already similar from change in behaviour associated with existing ties. A third process competes with both: two actors can converge on the same content because both follow a common source, without any tie between them or mutual influence. \citet[pp.~418--419]{mcphersonBirdsFeatherHomophily2001} situate similarity within the populations and settings that organise opportunities for contact, which means that a homogeneous network can arise through limited opportunity, selection among opportunities, or subsequent influence.

Two studies by the same author provide the corpus's sharpest contrast, and both model selection and shared content.

\citet[Table~1, pp.~1011--1016]{friemelOpinionLeadershipInfluence2015} reanalysed a four-wave panel of five newly formed school classes with 125 pupils, with conversation ties and television-programme preferences modelled as stochastic actor-oriented processes. Adding selection and programme co-nomination attenuated the influence-only estimates. In the influence-only model four of the five coefficients were significant at the 5\% level and the fifth at the 10\% level; in the full model none reached the 5\% level and one reached the 10\% level. The period-2 influence interactions were significant and opposite in direction in class A ($-.19$, SE $.09$) and class C ($.27$, SE $.10$). The full models therefore contain temporal heterogeneity rather than an absence of influence-related terms. Because the interactions estimate change from a reference period, the significance of a period-specific sum cannot be inferred without its covariance. Baseline selection receives support in one class, time interactions matter, and co-nomination receives support in all five; attenuation cannot be attributed to selection alone.

\citet[Table~3, pp.~318, 324--325]{friemelCoOrientationMediaUse2021} combined friendship nominations with media-content networks in three established grade levels covering 336 pupils. The models also include co-nomination and report positive influence coefficients for television and YouTube in all three grades.

The two studies differ on four conditions at once. The earlier observes television-related conversation in newly formed classes across four waves; the later observes friendship in established grade groups across three. The relationship type, the group stage, the content medium, and the treatment of time all differ. Selection may already be complete before observation begins in established groups, and co-nomination may absorb influence operating outside measured ties.

\subsection{Why the contrast is a result}

The temptation is to describe this as two studies disagreeing about whether influence occurs. That description is wrong in a way that matters. Neither study tests the other's conditions. The 2015 estimates are not evidence against influence in established friendship networks, and the 2021 estimates are not evidence against attenuation in newly formed classes. What the contrast establishes is narrower and more useful: controlling for selection and co-nomination does not make studies of different relationships and periods directly comparable, so the difference in estimates cannot be attributed to any of the four conditions that vary.

This illustrates \textbf{P3}: influence estimates are conditional on relationship type, group stage, and time specification, and cross-study comparison cannot attribute a difference to influence itself when these vary. The scope is longitudinal network models of media use that model selection and co-nomination. The claim would fail if a design holding these conditions constant still produced the observed contrast---which would mean the difference lies somewhere else.

The practical consequence is a design requirement. Measuring friendship and content-related conversation in the same groups, across formation and maintenance, with harmonised content measures and explicit time interactions, would separate relational from temporal explanations. The relevant comparison is a difference between estimated processes under shared observation conditions with uncertainty reported for that difference, not whether separate coefficients cross a threshold.

\section{The three propositions and their refutation conditions}
\label{sec:propositions}

The comparisons support three propositions. Each is stated so that it can be wrong.

\begin{description}[leftmargin=1.2em, style=nextline]
\item[P1 --- Task-dependence of relevance.]
For a given resource, judgements of relevance vary by task and criterion; recognition of expertise, attributed relationship importance, and reported source use are distinct outcomes and need not converge.
\emph{Scope:} settings in which actors purposively select sources.
\emph{Refutation:} a task-independent measure of resource value that predicts all three outcomes equally well across tasks.
\emph{Evidence:} \citet{palazzoloOrganizingInformationRetrieval2005}; \citet{taylorBuildingInterorganizationalRelationships2003}; \citet{johnsonMuchAdoNothing2018}.

\item[P2 --- Level-dependence of commonality.]
Commonality at one level of aggregation does not imply commonality at another.
\emph{Scope:} audience and media--public comparisons that measure both levels.
\emph{Refutation:} outlet overlap predicting article-level agreement, or issue correspondence predicting evaluative agreement, within the same participants.
\emph{Evidence:} \citet{websterDynamicsAudienceFragmentation2012}; \citet{rauchfleischTransnationalNewsSharing2020}; \citet{vuExploringWorldOutside2014}; \citet{vargoNetworkIssueAgendas2014}; \citet{hsiaoNetworkAgendaSetting2026}; \citet{luNetworkedFraming2024}.

\item[P3 --- Conditionality of influence estimates.]
Influence estimates are conditional on relationship type, group stage, and time specification, and cross-study comparison cannot attribute a difference in estimates to influence itself when these vary.
\emph{Scope:} longitudinal network models of media use that model selection and co-nomination.
\emph{Refutation:} a design holding these conditions constant that still produces the observed contrast.
\emph{Evidence:} \citet{friemelOpinionLeadershipInfluence2015}; \citet{friemelCoOrientationMediaUse2021}.
\end{description}

A fourth claim, about the literature rather than about communication, is stated with the bibliometric analysis in \appref{app:cocitation} as P4. It is numbered there and not here because it concerns how the field's references are cited, not how communication practices are compared. The propositions are related but not nested. P1 concerns what a resource is taken to be worth; P2 concerns the level at which similarity is measured; P3 concerns the conditions under which a process is estimated. Together they give the framework its content: a study's finding is comparable to another's when the problem, the resource, and the practice that the claim depends on are held fixed.

None of the reviewed studies supplies the refutation test for any proposition. That is not a weakness of the propositions but the reason they are worth stating: they specify what the corpus lacks, and \secref{sec:agenda} specifies the designs that would supply it.

\section{Boundaries: support, collective action, and representation}
\label{sec:extensions}

The three propositions describe what a network makes available and how that availability can be misread. They say nothing about whether the available resource is useful, whether participation is organised, or whether representation is equitable. Three bodies of work test those boundaries, and in each the framework's claim becomes narrower rather than failing.

\subsection{Availability, activation, and usefulness}

\citet{vriensDoesRiseInternet2018} followed 2,491 Dutch respondents who named core discussants in 2008--2010 and again in 2013. Frequent social-media users reported larger core networks with both more stable discussants and more turnover, while changes in media use were not significantly associated with changes in talking frequency. Continued availability with expanded membership fits reorganisation better than uniform withdrawal. With two waves, bounded nominations, and broad media-use measures, causal direction and channel-specific effects remain unresolved.

Availability is not activation. In a Korean immigrant church, \citet{leeRoleStatusDifferentials2019} distinguished everyday information from tangible help such as rides or childcare. Deacons had higher betweenness but not significantly higher degree in the information network. Tangible-help ties showed gender heterophily when family relations were included but no significant gender pattern after their removal. Because support ties are symmetrised, the study cannot identify providers, recipients, or assistance delivered. Organisational position therefore locates a member in a support structure without establishing what that member provides or obtains.

Usefulness is a third outcome. \citet{demetriadesDyingFitHow2024} separated network closure, reported vaccine discussion, and perceived descriptive and injunctive norms in a sample of 711 Illinois residents. Closure was weakly associated with lower knowledge ($r=-.08$) and did not distinguish vaccination status; indirect associations through discussion and descriptive norms linked closure with knowledge and behaviour. \citet{leeDiscussionNetworksResilience2022} distinguished network structure from discussion frequency and topical breadth among 599 undergraduates. Communication measures added explanatory information for perceptions of future and social competence, though some individual coefficients were marginal; separate topic models associated discussion of goals and achievements with perceptions of self, future, and social competence, and none of the examined topics predicted structured style.

Self-reported discussion networks behave similarly. \citet{moyPredictingDeliberativeConversation2006} found that network characteristics had mixed associations with conversational clarity, reasoning, and comprehension across two samples. More discussion cannot be equated with better deliberation.

The boundary is therefore a three-part distinction: retaining a partner, approaching that partner, and obtaining useful help are different outcomes, and no reviewed study follows all three in the same people. P1 concerns the first two. The third is outside its scope.

\subsection{Access, coordination, and collective reach}

Weak ties can extend access beyond close circles, and connective-action accounts explain how personalised expression and digital sharing organise participation \citep{granovetterStrengthWeakTies1973b,bennettLogicConnectiveAction2012}. The two explanations enter a study without doing the same work. In the Egyptian revolution study, weak ties and connective action supply contextual resources, structural holes define brokerage, and social-movement accounts inform the hypotheses \citep{abul-fottouhBrokerageRolesStrategic2018}. Liaison brokerage across ideological groups was proportionally more prevalent in the solidarity-period mention network than in the later schism-period network, though account sets also changed. In the later network, bloggers and nonparty activists had higher brokerage scores than their counterparts. Institutional affiliation and communication position therefore give different accounts of who connects participants.

Organisational contributions to public connection pose a related question. Across 48 English-language Islamic resistance organisation websites, generalist organisations supplied more hyperlinks and occupied broker and authority positions among generalists \citep{shumateConnectiveCollectiveAction2008}. The connective-public-good account concerns publicly accessible routes between sites rather than observed cooperation, and it predates the connective-action framework. Reading the two together distinguishes contributions to access from the organisation of participation.

Circulation supplies the sharpest case for separating a route from its content. \citet[pp.~1222--1226]{shinPoliticalRumoringTwitter2017} linked users to the rumours they posted about and projected those affiliations into a network of shared participation. Among believers the two detected communities corresponded exactly to anti-Obama and anti-Romney rumours; among rejecters the thirteen communities mixed the two targets, so shared participation did not sort by candidate once rejection was the common act. Of 782 tweets initially classified as rejecting a rumour because they cited a fact-checking site, 557 repeated the false claim while linking to the correction, and only 225 explicitly stated that it was false. A link to corrective information and a correction expressed in the message are different acts, and only the second is a statement a reader can act on.

\subsection{Visibility and representation}

Meaning comparisons require attention to who is represented. \citet{schultzStrategicFramingBP2012} constructed associative frames in BP crisis releases and news through co-occurrence and asymmetric conditional probabilities, finding that BP was associated with solutions in its own releases and that news partly echoed those associations while including additional political, legal, and protest actors. The presence of an actor does not establish its role; the interpretation depends on the surrounding associations, and aggregation over time leaves the direction of influence unresolved.

Source concentration and thematic diversity can coexist. \citet{veltriClimateChangeTwitter2017} found diverse climate themes alongside concentrated professional news sources among frequently shared links, drawing on social representations and media ecology. Diversity of discourse does not require diversity of sources. In \#COP26 Instagram posts, \citet{stoddartInstagramArenaClimate2025} found Indigenous themes visible while Indigenous actors remained at the periphery of the discourse network, with politicians and eco-celebrities receiving different evaluations.

Recognition can be genuine and bounded at the same time. \citet{kimRacializedBeautyVisibility2023} combined qualitative reading of 45 creator videos with a semantic network of 1,708 comments; the network located connected themes and guided close reading of how commenters described hurtful experiences, identification, and self-acceptance. Recognition occurred within commercial self-presentation, and an emphasis on monolid eyes may have narrowed the ethnic experiences represented; the 1,708 comments came from five of the 45 videos. Cross-role communication raises the same question: \citet{walterScientificNetworksTwitter2019} found scientists' cross-role mentions and retweets accompanied by differences in language, but such patterns do not establish sustained exchange or knowledge uptake.

The boundary is that structural position and common content are necessary but not sufficient for meaningful access or equitable representation. A source repertoire can reach professional expertise while omitting citizens' situated experience; an issue can be prominent while affected actors remain peripheral; recognition can be meaningful yet limited to particular identities.

\section{How networks are built: research practices across subfields}
\label{sec:practices}

The three comparisons drew on five constructions. Stated together they show what each makes observable and what it forecloses.

\paragraph{Nomination.} The consultation study and the Croatian campaign study each asked respondents to name partners; the nomination records a reported relationship, not observed contact \citep{palazzoloOrganizingInformationRetrieval2005,taylorBuildingInterorganizationalRelationships2003}. The correction-route coding in the rumour study \citep{shinPoliticalRumoringTwitter2017} is similarly a coder attribution, not a network edge, which is why the co-membership network and the correction measure must be read separately.

\paragraph{Trace.} The Twitter follow networks \citep{verweijTwitterLinksPoliticians2012,johnsonMuchAdoNothing2018} record a declared access relationship, not readership or use. The URL-sharing study \citep{rauchfleischTransnationalNewsSharing2020} constructs audience communities from sharing acts. \citet{bakshyExposureIdeologicallyDiverse2015a} distinguish the friend network (structural), the algorithmic ranking (presented), and the click (acted): each is a different edge, and the three do not collapse into one measure of exposure.

\paragraph{Affiliation, projected.} The audience duplication panel \citep{websterDynamicsAudienceFragmentation2012} connects outlets through shared measurement-period users. The \citet{barberaTweetingLeftRight2015} retweet and co-following networks connect users through shared behaviour toward issues. A projected edge does not record that the two parties knew one another or agreed.

\paragraph{Co-occurrence within text.} The issue and frame networks in the agenda and framing studies \citep{vargoNetworkIssueAgendas2014,schultzStrategicFramingBP2012,hsiaoNetworkAgendaSetting2026,luNetworkedFraming2024} record that two elements appeared in the same text unit; they do not record agreement, causation, or temporal order. The commenter network in \citet{luNetworkedFraming2024} connects users who commented on the same video, which is affiliation rather than co-occurrence; the two constructions coexist in that study because one tests whether text co-occurrence patterns vary with who gathers.

\paragraph{Longitudinal tie.} The two Friemel studies use stochastic actor-oriented models, which require a binary tie in each wave and distinguish selection from influence through simultaneous modelling of both processes \citep{friemelOpinionLeadershipInfluence2015,friemelCoOrientationMediaUse2021}. The result is sensitive to tie definition, period covered, and co-nomination inclusion---the conditions P3 names as the ones that cannot vary independently if findings are to be compared.

Several constructions in the corpus involve individuals who did not produce records for research purposes. Follower and co-following networks assign observed persons positions in a relational structure they cannot inspect. Co-mention and co-commenter graphs create recorded relationships between individuals who may never have interacted. Model-estimated ideological scores attribute political positions to individual users from behavioural traces. The reviewed studies relied on public-access data or platform APIs available at the time of collection, typically with pseudonymous or aggregated identifiers, and re-identification risks vary with data granularity. These costs are worth naming alongside what each construction makes observable, and designs that involve trace data at scale benefit from explicit justification alongside the methods description.

\appref{app:observation} lists what each construction does and does not establish. The list is not evaluative: constructions are appropriate or inappropriate for specific claims, not better or worse in general.

\section{What networks are asked to explain: theories and explanatory resources}
\label{sec:theories}

The reviewed applications invoke six theoretical families. Naming them clarifies what the explanatory resource dimension tracks and why it is not reducible to subfield or method.

\paragraph{Transactive memory and institutional sourcing.} Transactive memory predicts that recognised expertise, not held knowledge, drives consultation, so the relevant tie is nomination \citep{palazzoloOrganizingInformationRetrieval2005}. Institutional sourcing theory predicts that structural positions---reporter placements, interpretive community membership---shape source availability before any story is written \citep{tuchmanNewsNet1978,zelizerJournalistsInterpretiveCommunities1993}. Both are organisational, position-based mechanisms; they apply to different actors, timescales, and network types.

\paragraph{Audience-duplication and repertoire accounts.} The audience-duplication literature treats outlet co-use as an environment-level structural property \citep{websterDynamicsAudienceFragmentation2012}; repertoire accounts treat it as a person-level characteristic. The difference determines whether the network describes an environment or a behaviour, and what would count as a change.

\paragraph{Agenda setting and framing.} The agenda-setting tradition motivates co-occurrence networks: joint presentation can strengthen joint recall, making issue and attribute pairs the relevant units \citep{mccombsAGENDASETTINGFUNCTIONMASS1972,mccombsExpandedPerspectiveAgendaSetting2012}. Framing adds attributional structure \citep{entmanFramingClarificationFractured1993}. Both share the co-occurrence construction but differ in what co-occurrence is taken to do: activate salience or activate evaluation. Both operate as similarity-type mechanisms---tie strength indexes co-activation, not flow.

\paragraph{Homophily and selection--influence models.} Homophily provides the alternative explanation that ties track similarity through opportunity structure or selection rather than influence \citep{mcphersonBirdsFeatherHomophily2001}. Stochastic actor-oriented models decompose tie formation from behaviour change by treating both as simultaneous processes \citep{friemelOpinionLeadershipInfluence2015,friemelCoOrientationMediaUse2021}. The explanatory resource is a coordination mechanism: ties coordinate behaviour by transmitting a signal or by selecting for similarity.

\paragraph{Structural holes and connective action.} Brokerage theory predicts that actors spanning structural holes control information routes \citep{granovetterStrengthWeakTies1973b,abul-fottouhBrokerageRolesStrategic2018}. Connective action theory predicts that personalised sharing, not organisational coordination, organises digital participation \citep{bennettLogicConnectiveAction2012}. Both concern flow mechanisms, but differ in the scale and intentionality required.

\paragraph{Multitheoretical, multilevel models.} \citet{mongeTheoriesCommunicationNetworks2003} treat communication networks as explainable by multiple theories simultaneously across levels. The multitheoretical, multilevel framework specifies that a full account requires theory at the actor, dyadic, triadic, and global levels \citep{leeRoleStatusDifferentials2019}; it does not reduce to a single mechanism.

These six families are not exhaustive of communication network theory. What the list establishes is that the explanatory resource dimension is not reducible to the research practice or the communication problem: the same co-occurrence network can serve agenda-setting, framing, or shared-participant accounts, and the resource determines what the edge is taken to do. That is why comparing findings requires knowing not just what was measured but what the measure was invoked to explain.

\section{The literature's citation structure}
\label{sec:diagnostic}

This section turns from the applications to a bibliometric description of the same literature. The framework says when two findings may be compared; it does not say whether a field's citation structure is organised in a way that makes comparison likely. If interpreting a network requires substantive theory, a literature that routinely returns from representation to phenomenon should cite explanatory and methodological resources together at least as often as their prominence predicts. A conditional co-citation analysis of 2,103 records and 30 reference strings finds the opposite: method references are cited together more often than year-specific margins predict (ratio 1.36) and with theoretical references less often (0.82), while theory--theory and theory--empirical pairing sit at expectation. Both departures hold across sensitivity specifications, alternative type assignments, and leave-one-reference-out checks; two further departures in the full table do not and are reported as descriptive.

The pattern is consistent with a knowledge base in which methodological resources circulate as a shared toolkit while theory is cited within traditions. The analysis is cross-sectional, measures citation rather than adoption, and does not code citing records by communication problem, so it cannot show directly that problems fail to organise the literature. The design, the type-family table and figure, the robustness checks, and the statement of the proposition with its scope and refutation conditions (P4) are in \appref{app:cocitation}.

\section{Agenda: the designs that would test the propositions}
\label{sec:agenda}

Each proposition names the evidence that would settle it. This section specifies the designs in the order of the propositions, and adds the refutation test for the bibliometric claim.

\subsection{Following a resource across tasks}

P1 requires observing a resource's value across tasks rather than in one. In journalism, tracing monitoring, requests for information, corroboration, and editorial selection across assignments would show when source expertise, accessibility, or institutional standing enters a published account. Comparing publicly cited sources with those consulted in the background would distinguish visibility from use. A design combining following networks, communication logs, and content analysis would establish whether social-media monitoring precedes direct contact, and would test whether source-role specialisation holds across beats. The requirement is that the same source be followed through more than one task; P1 fails if a single criterion predicts its value in all of them.

\subsection{Measuring both levels on the same participants}

P2 requires the same people at two levels. Linking individual article selection, through clickstream or reading logs, with outlet-level repertoire descriptions would show whether within-outlet selection reproduces or qualifies between-community differences. Adding interpersonal discussion measures would test whether discussion amplifies, moderates, or reorients interpretations formed from exposure. For the media--public comparison, content-exposure tracking combined with pre-exposure elicitation of prior associations and post-exposure elicitation of how the same issues are organised would distinguish selective exposure, amplification, and reinterpretation. Tracking entry and exit in discussion threads would test whether aggregate divergence reflects turnover in who contributes. P2 fails if article-level agreement within shared outlets is as high as outlet overlap predicts, or if issue correspondence predicts evaluative agreement within individuals.

\subsection{Holding relationship and period constant}

P3 requires a design rather than a comparison. Recruiting participants at group formation and following them across at least four waves spanning 18--24 months, with separate elicitations of friendship and content-discussion ties, harmonised media-use measures, and stochastic actor-oriented models including selection, influence, and co-nomination terms for both tie types, would permit a test of whether influence parameters differ by relationship type and whether they change over time. The relevant statistic is a test of the difference between parameters with uncertainty reported for that difference, not separate significance tests. P3 fails if such a design reproduces the contrast observed between the two existing studies.

\subsection{Refuting the bibliometric claim}

P4, stated in \appref{app:cocitation}, requires a specification that fixes what the primary analysis leaves free. Three tests would be decisive. A specification conditioning on publication source and reference identity while preserving the year margins would remove the concern that venue or author overlap generates the pattern; the source-conditioned checks in \appref{app:cocitation} are a first step, and reference-identity consolidation remains to be done. Citation-context reading of the paired method references would test whether they serve shared explanatory commitments, which would reinterpret the over-pairing as a sign of integration rather than of tool-driven organisation. And a stratified analysis by communication problem, assigning citing records to the four problem families and asking whether problem predicts pairing after type is held constant, would test the organising claim directly. If problem predicts pairing as strongly as type does, P4 fails.

The corpus does not support the third test as constructed, because the problem assignment would be author-coded at the record level without independent verification. That limitation is the clearest reason to treat P4 as a stated proposition rather than an established finding, and the reason it is worth pursuing: it is testable with data this review already holds plus coding that a second coder could check.

\section{Limitations}

The corpus is bounded by Web of Science, the Communication category, the query, and available texts; coverage of 2026 ends on 31 August, so those records have shorter citation windows than earlier ones. The query and pilot records are preserved, but linking Q-R and the document-type filter to the final export relies on retrospective author confirmation, and the execution history has not been recovered.

Three limits bear on the substantive comparisons. Purposive selection targets contrast, so the corpus over-represents studies whose findings do not agree, and a different selection would foreground other tensions. Full-text access was uneven, and abstract-level evidence is identified wherever it is used. No second coder was available, so selection and coding rest on one reader, and the LLM assistance documented in \secref{sec:corpus} is not independent verification.

Coverage is uneven by subfield. Newsroom collaboration, support across life stages, and cultural and identity research receive less treatment than the agenda and audience literatures, and the agenda comparisons cover selected United States, Chinese, and Turkish settings rather than a general history of either tradition. Opinion leadership rests on a two-study comparison. Platform-address studies observe address rather than sustained exchange.

The bibliographic analysis carries the limits stated in \appref{app:cocitation}: reference strings are not consolidated works, type assignment is author-coded without a reliability estimate, the conditional baseline does not control for topic, and source conditioning is limited to the selected pairs; co-citation records shared citation rather than adoption. Finite simulation does not yield uniform rejection decisions across specifications, which is why each departing combination is reported together with its robustness rather than summarised.

\section{Conclusion}

Communication network research shares a vocabulary and does not share its objects. Consultation, media use, citation, and lexical association are all represented as edges, and a reader can move between studies as though the edge meant the same thing. This review argued that it does not, and offered a framework for judging when two findings may be placed side by side: same problem, same explanatory resource, and practices that make the same observation. The framework formalises that rule through the distinction between abstracting a phenomenon into a network and representing that network in data, and it locates the point at which a comparison stops being licensed.

Three propositions follow, each with a scope and a refutation condition. Resource relevance is task- and criterion-dependent. Commonality at one level of aggregation does not imply commonality at another. Influence estimates are conditional on relationship type, group stage, and time. None of the reviewed studies supplies the refutation test for any of them, and that absence is itself informative: the corpus contains many studies that measure a resource, a level, or a condition, and few that hold the others fixed while varying it.

The bibliometric section describes the literature's citation structure, and the pattern bears on all three propositions. Citation pairing follows resource type. Method references are cited together more than their margins predict, and theory with method less; both departures hold across sensitivity specifications, alternative type assignments, and leave-one-reference-out checks, while two further departures are reported as descriptive. Locally, particular theoretical, empirical, and methodological works are co-cited far above expectation within their traditions, while theory--theory and theory--empirical pairing sit at expectation in aggregate. A field of this shape can accumulate procedures quickly, because procedures are shared, while its explanations remain local, because the theory that would license comparison does not travel with the methods.

The bibliometric pattern is consistent with the framework's motivation, and that consistency is the review's interpretative claim rather than a demonstrated mechanism. A knowledge base organised by what its studies use rather than by what they try to explain will admit studies that share a method without sharing the commitments that would make their findings commensurable. If that reading is right, comparing measures does not produce comparison; that is also why the three propositions are stated as conditions rather than as findings. It is also why the framework is worth applying before findings are pooled: the question is not whether two studies measured the same thing, but whether their measurements bear on the same claim.

\FloatBarrier
\printbibliography[title={References}]

\appendix

\section{Retrieval and provenance}
\label{app:retrieval}

\subsection{Query and export}

The retrieval combined the Web of Science Core Collection Communication category with network-analysis expressions. The expression below is query Q-R from the archived search registry. On 19 September 2026 the author confirmed that it was used, with the Article/Review document-type filter, for the final export. That confirmation is retrospective: the pilot records document no export at that stage, and the final query-execution history has not been recovered. The pilot returned 2,181 hits; after the document-type filter and deduplication the export contains 2,114 unique records. The archived records do not allow the 67-record difference to be reconstructed. Line breaks are for readability.

\begin{quote}
\small\ttfamily\raggedright
WC=("Communication") AND TS=(("network*" NEAR/3 analy*) OR
"two-mode network*" OR "two mode network*" OR "bipartite network*" OR
"multiplex network*" OR "temporal network*" OR "dynamic network*" OR
"exponential random graph*" OR "stochastic actor-oriented model*" OR
"relational event model*" OR "quadratic assignment procedure" OR
"network autocorrelation" OR (network* NEAR/5 centralit*) OR
(network* NEAR/5 betweenness) OR (network* NEAR/5 modularit*))
\par
\end{quote}

The document-type filter is reported separately because it is not part of the archived query string.

Five Plain Text exports dated 31 August 2026 yield 2,114 unique records after deduplication by identifier. Document types are preserved as exported: 1,905 \emph{Article}, 28 \emph{Review}, and composite types including \emph{Article; Early Access}.

\subsection{Vocabulary fields}

Author keywords (DE) and Keywords Plus (ID) are present in 1,419 records; their union covers 2,042 records, while title-and-abstract lexical analysis retains all 2,114. These measures describe vocabulary coverage, not theoretical importance. The fields have different origins: DE records author-supplied terms, ID records indexing terms assigned by the database, and a study retrieved through ID may not use the term in its own description.

\subsection{Pair-eligible records}

The primary co-citation view uses the \NetworkEligibleRecords{} records carrying at least two distinct nonempty reference strings. Ten records carry no cited references and one carries a single reference; neither can contribute a co-citation pair. Cited-reference counts reported in the source fields cannot recover reference strings that were absent from the export.

\begin{table}[htbp]
\caption{Threshold-induced co-citation views. Reference frequency is counted within the \NetworkEligibleRecords{} records with at least two distinct references. Edges are distinct pairs with positive co-citation.}
\label{tab:thresholds}
\centering
\normalsize
\renewcommand{\arraystretch}{1.12}
\begin{tabular}{@{}rrrrr@{}}
\toprule
Minimum frequency & Identities & Edges & Records citing & Records co-citing \\
\midrule
30 & 60 & 1,172 & 1,245 & 797 \\
40 & 30 & 329 & 1,101 & 563 \\
50 & 15 & 95 & 899 & 346 \\
\bottomrule
\end{tabular}
\end{table}

\section{Co-citation analysis: specification}
\label{app:cocitation}

\subsection{The implication to be tested}

The framework says when findings can be compared. It does not say whether a field's knowledge base is organised in a way that makes comparison likely. The network-science argument on which the framework rests gives that question a testable form. If interpreting a network requires substantive theory, then a literature that routinely makes the return from representation to phenomenon should cite explanatory and methodological resources together at least as often as their prominence predicts. If instead methods circulate as a toolkit detached from the explanations that license them, method references should be paired with one another above expectation and with theoretical references below it. The analysis reported here is a first test of that contrast.

The test is indirect in one respect that matters. It classifies cited references by the kind of resource they supply; it does not assign citing records to communication problems. It can therefore show whether theory and method are cited together, but not whether studies addressing the same problem share references. The second question is the stronger form of the claim, and \secref{sec:agenda} specifies the coding that would test it.

\subsection{Data and method}

The test uses the review's candidate corpus: \NetworkEligibleRecords{} records carrying at least two distinct reference strings. The \NetworkPrimaryNodes{} reference strings cited by at least 40 records form the primary view; thresholds of 30 and 50 give broader and narrower descriptive views (\tabref{tab:thresholds}). All 435 possible pairs, including the 106 with zero co-citation, enter the comparison. In this view 1,101 of the 2,103 pair-eligible records cite at least one of the 30 strings, 1,095 are exchangeable under the trade procedure, and 563 cite at least two, contributing 1,445 record-level co-citation incidences; the remaining 1,002 records contribute no pair. The 30 strings comprise 12 method or software, 9 conceptual or theoretical, 5 empirical, and 4 mixed or uncertain references, giving 66 possible method--method pairs and 108 possible theory--method pairs.

Raw co-citation counts confound association with prominence: two works may be cited together often simply because each is widely cited. Co-citation was introduced as a measure of intellectual association between documents \citep{smallCocitationScientificLiterature1973}, and the dependence of raw counts on citation prominence is a standard limitation of the measure in bibliometric reviews \citep{zupicBibliometricMethodsManagement2015}. Dividing by marginal totals rescales the counts but does not condition on the full incidence structure, and similarity measures that are not monotonically related need not order pairs identically \citep{vaneckHowNormalizeCooccurrence2009}. Mapping software applies such normalisation to co-occurrence data before laying out a map \citep{vaneckSoftwareSurveyVOSviewer2010}, so the measure chosen changes what a map displays. The analysis here therefore compares observed co-citation with a conditional expectation obtained by fixed-margin randomisation, the Curveball procedure \citep{stronaFastUnbiasedProcedure2014}, for which \citet{verhelstEfficientMCMCAlgorithm2008} had earlier given an equivalent formulation with a different implementation: reference incidences are traded between records within the same citing year, preserving each record's number of selected strings and each string's citing-record frequency within that year. Fixed-margin randomisation is an established null for extracting structure from bipartite projections \citep{nealBackboneBipartiteProjections2014,nealComparingAlternativesFixed2021}. Four chains of 10,000 saved states were run; convergence and the full specification are reported in the subsections below. Multiplicity is controlled separately within two families, the 435 pairs and the ten resource-type combinations, both with Benjamini--Yekutieli adjustment \citep{benjaminiControlFalseDiscovery2001}; the separation matters because the ten combinations are not independent of the 435 pairs.

Reference strings were assigned to four resource types---conceptual or theoretical, empirical, method or software, and mixed or uncertain---on the basis of the cited resource's role rather than the citing paper's topic. Assignments follow the reference list and are recorded with the label boundaries in Supplementary Data S1, which also records the strings that could not be assigned confidently.

\subsection{Result}

\begin{figure}[htbp]
\centering
\includegraphics[width=0.92\textwidth]{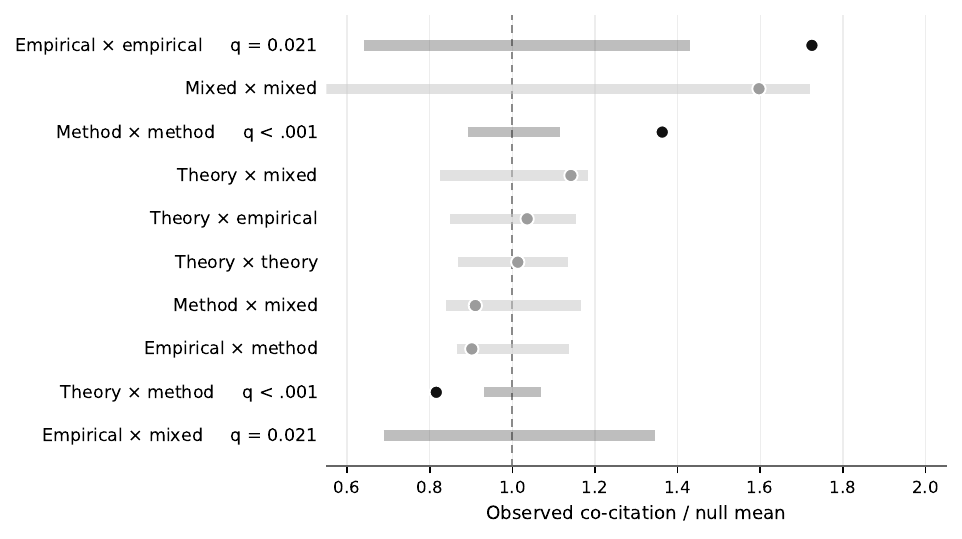}
\caption{Observed co-citation relative to conditional expectation, by reference-type family. Values above one indicate pairs cited together more often than the year-specific margins predict. Filled points mark type combinations with Benjamini--Yekutieli adjusted $q<.05$; open points are not distinguishable from the conditional expectation. Bars span the 2.5th to 97.5th percentile of the exact null distribution, obtained by summing the saved chain draws within each type combination.}
\label{fig:types}
\end{figure}

The type-family comparison yields a clear pattern (\figref{fig:types}, \tabref{tab:types}). Citation pairing follows resource type.

\begin{table}[htbp]
\caption{Co-citation by reference-type family, primary specification. Observed counts are record-level co-citation incidences; expectations preserve citing year and the selected-reference margins. $q$ values are Benjamini--Yekutieli adjusted within the family of ten type combinations.}
\label{tab:types}
\centering
\normalsize
\setlength{\tabcolsep}{4pt}
\renewcommand{\arraystretch}{1.10}
\begin{tabular}{@{}lrrrr@{}}
\toprule
Type pair & Observed & Expected & Ratio & $q_{BY}$ \\
\midrule
Method $\times$ method & 324 & 237.76 & \textbf{1.36} & \textbf{$<.001$} \\
Theory $\times$ method & 380 & 465.78 & \textbf{0.82} & \textbf{$<.001$} \\
Empirical $\times$ empirical & 35 & 20.29 & \textbf{1.73} & \textbf{.021} \\
Empirical $\times$ mixed & 17 & 31.97 & \textbf{0.53} & \textbf{.021} \\
\midrule
Theory $\times$ theory & 176 & 173.72 & 1.01 & 1.00 \\
Theory $\times$ empirical & 139 & 134.21 & 1.04 & 1.00 \\
Theory $\times$ mixed & 112 & 98.09 & 1.14 & .69 \\
Empirical $\times$ method & 146 & 161.89 & 0.90 & .72 \\
Method $\times$ mixed & 103 & 113.14 & 0.91 & 1.00 \\
Mixed $\times$ mixed & 13 & 8.14 & 1.60 & .69 \\
\bottomrule
\end{tabular}
\end{table}

Four combinations depart from expectation. Method--method pairs are over-represented (324 observed against 237.76 expected, ratio 1.36). Theory--method pairs are under-represented (380 against 465.78, ratio 0.82), and the two departures are complementary: the same records that pull method references together omit theoretical reference strings that the margins would predict. Empirical--empirical pairs are over-represented (35 against 20.29, ratio 1.73), and empirical--mixed pairs are under-represented (17 against 31.97, ratio 0.53). The remaining six combinations are indistinguishable from expectation, including theory--theory (ratio 1.01) and theory--empirical (ratio 1.04), which are the combinations that would carry evidence of shared explanatory commitment across problems.

The four significant combinations are not equally robust across the four sensitivity specifications reported below (\tabref{tab:sensitivity}). Method--method and theory--method hold under all four, with ratios between 1.20 and 1.42 and between 0.77 and 0.82. Empirical--empirical also holds under all four (1.71--1.77). Empirical--mixed holds under three, reaching $q=.071$ at the threshold-30 view and $q=.056$ when 2026 records are excluded, which indicates that the departure rests on a small number of low-frequency pairs.

Beyond these specifications, the principal departures are robust to how the strings are typed and to which strings are in the view (\tabref{tab:robustness}). The four departing combinations were recomputed under four alternative labellings of strings whose type is arguable: the three classic empirical studies (McCombs, Katz, and Gamson) reclassified as theory; the four mixed or uncertain strings reclassified as theory; both changes together; and Monge and Contractor reclassified as method. Method--method and theory--method keep their direction and significance under every relabelling, with ratios between 1.33 and 1.36 and between 0.78 and 0.85. Empirical--empirical and empirical--mixed do not: moving the classics to theory leaves empirical--empirical resting on a single pair of strings, at a ratio of 7.32, and removes the empirical--mixed combination altogether, while moving the mixed strings to theory removes it as well.

Each of the 30 strings was then deleted in turn, with the combinations recomputed against the full-matrix baseline. Method--method ranges from 1.27 to 1.44 and theory--method from 0.75 to 0.88; every leave-one-out estimate for these two combinations is significant at $p<.01$. Empirical--empirical ranges from 0.91 to 2.51 and empirical--mixed from 0.23 to 0.71, and the empirical--empirical estimate falls below one when Barber\'{a} and Colleoni are removed. Those two departures therefore rest on a small number of strings and should be read as descriptive.

Deleting a single string also locates the method--method excess. The five largest pairwise excesses are Blondel--Bastian, Bastian--Jacomy, Blondel--Jacomy, Blei--Maier, and Freeman--Wasserman (1995): community-detection and topic-modelling software, together with two general methods texts. The excess is a cluster of computational tools cited as a set rather than methodological work distributed across the field. The theory--method deficit is correspondingly largest for Granovetter's weak-ties paper, whose removal raises the ratio from 0.82 to 0.88.

\begin{table}[htbp]
\caption{Co-citation ratios under alternative type assignments and leave-one-string-out. Entries are observed/expected ratios; the primary specification is in \tabref{tab:types}. Rows that relabel strings change the composition of the families as well as their counts, so ratios are not strictly comparable across rows.}
\label{tab:robustness}
\centering
\small
\setlength{\tabcolsep}{4pt}
\renewcommand{\arraystretch}{1.12}
\begin{tabular}{@{}lrrrr@{}}
\toprule
View & Method$\times$method & Theory$\times$method & Empirical$\times$emp. & Empirical$\times$mixed \\
\midrule
Primary & 1.36 & 0.82 & 1.73 & 0.53 \\
Classics as theory & 1.36 & 0.84 & 7.32$^{a}$ & 0.24 \\
Mixed as theory & 1.36 & 0.83 & 1.73 & --- \\
Both relabellings & 1.36 & 0.85 & 7.32$^{a}$ & --- \\
Monge as method & 1.33 & 0.78 & 1.73 & 0.53 \\
\midrule
Leave-one-out range & 1.27--1.44 & 0.75--0.88 & 0.91--2.51 & 0.23--0.71 \\
\bottomrule
\end{tabular}

\vspace{2pt}
{\footnotesize $^{a}$~Single remaining pair of strings. ---~No pair of strings carries this combination.}
\end{table}

\subsection{What the analysis does and does not show}

The pattern is consistent with a knowledge base in which methodological resources circulate as a shared toolkit. Method references are cited together, the theory references that would license interpreting them are cited apart from them, and the excess is concentrated in a few computational tools rather than spread across the field's methodological work. The interpretation rests on an inference that the data do not directly observe: a literature that routinely returned from representation to phenomenon would be expected to cite theory alongside the methods it interprets, and that pairing is what falls below expectation. Because citing records are not assigned to communication problems, the analysis cannot show that problems fail to organise the literature; it shows that the theory--method conjunction is under-represented relative to prominence.

The design is cross-sectional. It describes the structure of the knowledge base at one point, not the rate at which methods or theories accumulate, and it cannot show whether the asymmetry is stable, widening, or narrowing.

Three limits qualify the inference. The analysis describes citation organisation; it does not measure theoretical integration.

First, co-citation records shared reference use, not adoption. Two studies can cite the same theoretical work for different reasons, including as a foil, and two studies can share a commitment without sharing a citation. The analysis measures bibliographic proximity, and the substantive interpretation requires citation-context reading.

Second, the type assignment is the author's. Moving a string between theory and method changes both cells that carry the principal result. The relabelling and leave-one-out checks in \tabref{tab:robustness} bound the effect of the relabellings considered: method--method and theory--method survive all of them, while empirical--empirical and empirical--mixed do not and are reported as descriptive. The checks substitute for, but do not replace, independent coding; no reliability estimate exists for the assignment.

Third, the design fixes margins but not composition. The conditional baseline does not control for topic, publication source, or author overlap. Source-conditioned checks on the selected pairs preserve their positive excess under every one of the 278 source deletions, and three of the four retain adjusted evidence under a year-and-source conditioned model, but neither specification resolves retrieval or identity uncertainty. The subsections below report those results and the identity checks.

A fourth consideration is interpretive rather than technical. The aggregate theory--method deficit is compatible with at least two accounts: the field's theory is thin, or theory is carried by tradition-specific canons whose citation does not track method use, which would produce the deficit even in a theoretically mature field. The selected pairs show that citation is locally organised around recognised conjunctions---a framing theory with framing research, a topic model with its communication applications, homophily with polarisation research---but none of the four is a theory--method pair, and in aggregate the theory--theory and theory--empirical combinations sit at expectation. The local evidence therefore does not support the second account in its strong form. What it suggests is narrower: where theory is cited heavily, it is cited within a tradition, and the methods that travel widely are not cited with it.

Read together, the aggregate and the local results describe a literature with a shared methodological toolkit and tradition-specific theory. Method references are cited together, and the excess sits in a cluster of computational tools that travel as a set. Theory is not cited together more than chance, nor with method; where particular theoretical works are heavily co-cited, they are co-cited with the empirical or theoretical work of their own tradition (\tabref{tab:pairs}). A field of this shape can accumulate procedures quickly, because procedures are shared, while its explanations remain local, because the theory that would license comparison does not travel with the methods.

\subsection{A statement that can be refuted}

The analysis supports a fourth proposition, stated here so that it can fail.

\begin{description}[leftmargin=1.2em, style=nextline]
\item[P4 --- Resource-typed citation structure.]
In the candidate literature, citation pairing follows the type of resource: method references are paired with one another more than their margins predict, and theoretical references are paired with methodological ones less.
\emph{Scope:} the \NetworkEligibleRecords{} pair-eligible records, of which 563 co-cite at least two of the 30 strings cited by 40 or more, in the views reported here.
\emph{Refutation:} a specification that fixes year, publication source, and reference identity and finds theory--method pairing at or above expectation, or method--method pairing at or below it; citation-context reading showing that co-cited method references are cited for shared explanatory commitments; or a problem-level coding of citing records in which shared communication problem predicts pairing at least as strongly as shared resource type.
\emph{Evidence:} \tabref{tab:types}, \tabref{tab:robustness}, \tabref{tab:pairs}, \tabref{tab:sensitivity}; the subsections below.
\end{description}

\subsection{Randomisation}

The randomisation is the Curveball algorithm \citep{stronaFastUnbiasedProcedure2014}, which trades entries between records while preserving row and column totals and samples matrices with fixed margins. An equivalent formulation, with a different implementation, was published earlier by \citet{verhelstEfficientMCMCAlgorithm2008}; the 2016 corrigendum to the published Curveball description acknowledges that earlier work, and a 2015 corrigendum corrects the supplementary software listing. Neither correction concerns the procedure used here. The trade procedure preserves row and column totals and allows self-transitions, which means the row constraint concerns selected strings rather than the entire bibliography. Each sweep attempts as many trades as there are exchangeable records; 1,095 rows were tradable in the primary view, distributed across 30 strata defined by citing year. Margins were preserved in all four chains. The fraction of trades that changed the incidence matrix was approximately 0.634 within each chain, which indicates that the sampler moved rather than proposing and rejecting.

Chains start from the observed matrix with seeds 20260915 + 104729$c$ for $c=0,1,2,3$, 50 burn-in sweeps, and ten sweeps between saved states, retaining 10,000 states per chain. Convergence was assessed with the split $\widehat{R}$ and effective sample size statistics: the primary edge statistics have a maximum split $\widehat{R}=1.00028$ and a minimum effective sample size of 36,747 of 40,000; the type-family statistics have a maximum split $\widehat{R}=1.000098$ and a minimum effective sample size of 38,133. Trace, cumulative-mean, and autocorrelation plots for the selected pairs and for the extreme pairs are held with the archived diagnostics.

\subsection{Test statistics}

The primary test compares each observed pair count with the distribution generated under the conditional baseline. Because 435 pairs are tested, and because the finite simulation produces a discrete null distribution, the smallest attainable two-sided probability is $2/(B+1)$ for $B$ retained states; the reported $q$ values are therefore bounded below by that resolution. Upper-bound Monte Carlo adjustment was applied as a check and is reported with the primary results. The type-family comparison tests ten combinations in a separate family.

\subsection{Selected pairs}

Four pairs were selected contrastively, before the conditional analysis, to represent different conjunctions of conceptual and methodological resources: Entman and Gamson for framing, Blei and Maier for topic modelling, Barber\'{a} and Colleoni for homogeneity and polarisation, and Fraser and Habermas for publics. Their selection from the same candidate pool that generated the baseline makes the analysis exploratory, and the $q$ values are reported within the 435-pair family rather than separately.

\begin{table}[htbp]
\caption{Contrastively selected reference pairs. Observed co-citation is compared with the conditional expectation; the ratio is the excess over that expectation. Contexts are drawn from abstract review and describe what the pairing indexes, not what either work contributes to a citing argument.}
\label{tab:pairs}
\centering
\small
\setlength{\tabcolsep}{4pt}
\renewcommand{\arraystretch}{1.12}
\begin{tabular}{@{}llrrrl@{}}
\toprule
Pair & Context & Obs. & Exp. & Ratio & $q_{BY}$ \\
\midrule
Entman--Gamson & Framing, meaning, contested issues & 24 & 3.14 & 7.64 & .012 \\
Barber\'{a}--Colleoni & Homogeneity, polarisation, echo chambers & 18 & 2.46 & 7.32 & .012 \\
Blei--Maier & Topic modelling across communication problems & 17 & 2.59 & 6.55 & .012 \\
Fraser--Habermas & Public sphere, counterpublics, online discussion & 10 & 1.53 & 6.55 & .012 \\
\bottomrule
\end{tabular}
\end{table}

\subsection{Sensitivity specifications}

Four alternatives use four chains of 5,000 saved states each: lowering the reference threshold to 30, excluding 2026 records, excluding the flagged reference string, and retaining only records whose reported and exported reference counts agree. Record-deletion specifications retain the primary node set. The selected-pair excesses and the directions of the method--theory and method--method contrasts persist across all four; the empirical--mixed departure does not, as reported in \tabref{tab:sensitivity}.

Reference identity was checked separately. Exact-DOI consolidation reduces the node count from 60 to 59 at the frequency-30 view and leaves the frequency-40 view at 29 identities, with one string reclassified; a title-candidate scenario reduces it further. The two Wasserman strings that refer to the same book are retained separately in the frozen view, and the review reports this rather than correcting it, because correcting identity after inspection would change the view under test. As a check, merging the two strings into a single 29-identity view (union per record) produces method--method 1.40 ($q_{BY}<.001$) and theory--method 0.81 ($q_{BY}<.001$), confirming that the two principal departures do not depend on counting this book twice.

\subsection{Source conditioning}

Publication-source checks recompute co-citation after deleting each of 278 candidate publication sources in turn. Each of the five archived key pairs---the four in \tabref{tab:pairs} plus Entman--Maier---retains a positive co-citation range under every deletion, which addresses the concern that a single venue generates a pairing. A year-and-source conditioned model fixes different margins; three of the four tabled pairs retain adjusted evidence under it, and Fraser--Habermas does not at the .05 level. The two models are complementary rather than alternative tests, because they condition on different quantities.

\begin{table}[htbp]
\caption{Sensitivity of the four departing type combinations. Each alternative fixes a different margin or excludes a subset; the primary specification is in \tabref{tab:types}. Entries give the observed/expected ratio with the adjusted $q$ in parentheses.}
\label{tab:sensitivity}
\centering
\small
\setlength{\tabcolsep}{4pt}
\renewcommand{\arraystretch}{1.12}
\begin{tabular}{@{}lrrrrr@{}}
\toprule
Specification & Method$\times$method & Theory$\times$method & Empirical$\times$emp. & Empirical$\times$mixed & $q<.05$ \\
\midrule
Primary (40) & 1.36 ($<$.001) & 0.82 ($<$.001) & 1.73 (.021) & 0.53 (.021) & 4 \\
Threshold 30 & 1.20 ($<$.001) & 0.77 ($<$.001) & 1.77 ($<$.001) & 1.16 (.071) & 3 \\
No 2026 records & 1.36 (.001) & 0.82 (.001) & 1.76 (.016) & 0.55 (.056) & 2 \\
No flagged string & 1.42 (.001) & 0.81 (.001) & 1.71 (.020) & 0.52 (.020) & 4 \\
Reference-count match & 1.37 (.001) & 0.80 (.001) & 1.77 (.018) & 0.52 (.037) & 4 \\
\bottomrule
\end{tabular}
\end{table}

\section{Case roster}
\label{app:roster}

The 31 applications are recorded in Supplementary Data S1 with role, comparison assignment, inclusion rationale, evidence level, and source locator. Thirteen carry the principal comparisons, six contextualise or qualify them, and 12 test boundaries. One further application, on a televised political debate, is reported below and is not counted toward the 31.

\citet{fuhseAnalyzingNetworksCommunication2023} examines a 100-minute televised debate among six party representatives in Germany in 2012. Supportive and adversarial interruptions, together with accounts of other parties' actions, reveal different forms of engagement. The Social Democrats and Greens support one another; Christian Democrats and Free Democrats direct moderate attacks at one another; the Left and Pirate parties receive and initiate comparatively little engagement in the selected event types. These patterns broadly correspond to governing, established-opposition, and outsider roles. The finding concerns relationships presented in this debate, not an independently established effect of institutional roles or audience responses.

Roles describe a study's function in this synthesis, not its quality. The same study could carry a different role in a different comparison, and several studies appear in more than one section because they bear on more than one proposition.

\section{Observation guide}
\label{app:observation}

\subsection{What each construction makes observable}

\tabref{tab:guide} lists the constructions that recur in the corpus and states, for each, what the observation does not establish. The list is the operational form of the comparability rule: two findings can be juxtaposed only when their constructions support the same kind of claim.

\begin{table}[htbp]
\caption{Edge constructions and their warrants. The final column states what a finding from each construction cannot establish.}
\label{tab:guide}
\centering
\small
\setlength{\tabcolsep}{4pt}
\renewcommand{\arraystretch}{1.14}
\begin{tabular}{@{}p{0.20\textwidth}p{0.30\textwidth}p{0.42\textwidth}@{}}
\toprule
Construction & What is observed & What the observation does not establish \\
\midrule
Nomination & An actor names another as a partner, expert, or source & That advice was sought or taken; that the nominee was consulted \\
Affiliation, unprojected & Actors co-participate in an event, outlet, or category & That participants know or influence one another \\
Affiliation, projected & Two outlets or sources share users or are co-mentioned & Direct contact, cooperation, or shared editorial stance \\
Co-occurrence within text & Two elements appear in the same unit & Agreement, endorsement, causal linkage, or temporal order \\
Longitudinal tie & Who names whom at successive occasions & Causal direction between tie and behaviour without a model \\
Trace & A logged act such as a click, share, or follow & Attention, comprehension, or endorsement \\
\bottomrule
\end{tabular}
\end{table}

\subsection{Questions to apply to each comparison}

For each comparison the review asked: what is the communication problem; which explanatory resource does the study invoke; what does the construction make observable; what population and period; what would the study have found if the proposed explanation were wrong; and what would a different construction have shown. The last two questions are the ones that most often distinguish a comparison from a summary, and they are the reason several paragraphs in \secref{sec:relevance} to \secref{sec:influence} state what the study could not establish alongside what it found.

\section{Code and data availability}

The analysis and verification code is supplied in Supplementary Code S1. Its README identifies modules, input requirements, and the limits of reproduction without the licensed corpus. Archived numerical summaries allow the reported comparisons to be checked; recreating the original export requires authorised database access and cannot recover missing historical search settings.

\section*{Disclosure of AI assistance}

Large language models assisted literature screening and coding, comparison of source texts, and the development and checking of analysis and visualisation code. The author reviewed all substantive claims against source texts, and the verification record identifies which claims were checked at page level and which rest on abstracts. The author retains responsibility for the manuscript's content.

\end{document}